\documentstyle[11pt,paspconf,epsf]{article}

\begin{document}

\title{First results from the EROS-II microlensing experiment}

\author{Olivier Perdereau\altaffilmark{1}}
\affil{Laboratoire de l'Acc\'el\'erateur Lin\'eaire\\
IN2P3-CNRS \& Univ. de Paris-XI\\
BP 34 F-91898 Orsay Cedex, France}

\altaffiltext{1}{\bf on behalf of the EROS-II Collaboration}

\begin{abstract}

EROS-II  
is a second generation microlensing experiment. 
The experimental setup, in operation
at the European Southern Observatory (ESO) at La Silla (Chile) 
since mid-1996 is briefly described together with its scientific objectives. 
The first results from our microlensing searches  
towards the Small Magellanic Cloud (SMC) and 
the Galactic Plane are presented.  
We also give some results from a dedicated campaign which took place in 96-97, and  
aimed at studying   
magellanic Cepheids systematically. We conclude  
by an over\-view of the semi-auto\-mated supernov\ae\ 
search. 
\end{abstract}


\section{Introduction}

The microlensing effect was proposed ten years ago (Paczynski 1986) as a 
unique experimental signature of MACHOs. 
Dark compact baryonic objects (MACHOs) are plausible components of the galactic dark matter.  
The flux of an observed star is gravitationally deflected if one of these 
objects passes close to the line of sight. 
Because the image distorsion is undetectable, 
one is left with a transcient 
magnification of the total flux. 
Under the simplifying assumption of a point-like deflector in uniform motion 
and static point-like source and observer, the time variation of the apparent source luminosity 
has a universal, time-symmetric and achromatic shape. 
Its time scale $\Delta  t$, defined as the ratio between the deflector's transverse speed 
($v_T$) and a 
characteristic length of the phenomenon, 
the Einstein radius, is the only observable carrying 
physical information. It may be expressed as 
\begin{displaymath}
\Delta t (days) \ =\ 39\ \left( \frac{v_T}{100 km.s^{-1}}\right)^{-1}\ 
\sqrt{\frac{M}{M_{\odot}}}\ \sqrt{\frac{L}{10\ kpc}}\ \sqrt{x(1-x)}
\end{displaymath}
where $L$ is the distance to the source, $xL$ the distance to the deflector and 
$M$ its mass. The observable characterizing the event rate is the optical depth, 
denoted $\tau$, defined as the probability for observing a star being amplified by 
a factor greater than or equal to 1.34.
\par
This approximation may be violated in several ways, most of them being useful 
in breaking (some of) the degeneracy on the inferred physical or geome\-trical 
parameters extracted from the events. Among these, the deflector may be binary, 
leading to singularities in the amplification known as caustic crossing.
The Earth is rotating around the Sun ; the source or the deflector 
may be orbiting as well. These orbital motions may distort 
the light curve, leading to observable effects known as parallax and ``xallarap'' 
in the first 2 cases. These are discussed in more details in D. Bennett 1999. 
\par
Few years after B. Paczynski's proposal,  
few microlensing amplification of stars were 
observed in two different directions, the LMC (Alcock 1993, Aubourg 1993) 
and the Galactic Bulge (Udalski 1993). 
This field has now entered a more quantitative era. EROS-I has 
isolated 2 candidates over 3 years of running (Ansari 1996).
The Macho collaboration 
has taken data from 1993 until now and has a handful microlensing candidates 
towards LMC, and several hundreds towards the Bulge (Alcock 1999). An upgraded 
version of OGLE is now running, and gave about 40 alerts towards the Bulge 
(Udalski 1997).
\par
\begin{figure}
\begin{center}
\plotfiddle{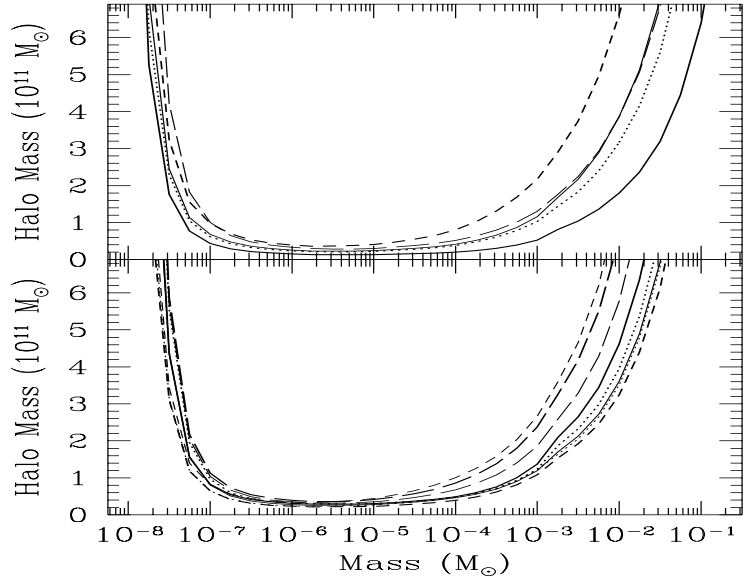}{8 true cm}{0}{65}{40}{-230}{-60}
\caption{\label{ermac}Combined EROS and  Macho upper limit on the 
halo mass fraction made of compact objects, as a function of their mass. 
Different line styles correspond to different models ; the mass of a ``standard'' 
isothermal, spherical halo responsible for the Galactic rotation curve 
is $4\times 10^{11}M_{\odot}$.}
\end{center}
\end{figure}
The observed optical depth towards LMC indicate a total halo MACHO mass fraction  
within a factor of two from the total 
required to explain the Galactic rotation curve. 
The combined null result from the search for short duration 
events by EROS-I and Macho exclude a significant contribution 
of objects in the mass interval [$10^{-7}$, $10^{-3}$]$M_{\odot}$, as shown on figure 
\ref{ermac}(Alcock 1998). 
In addition the rather long time scales associated with the observed  events indicate  
large lens masses, which is difficult to accommodate with known stellar populations.  
\par 
To address these questions, EROS started as early as 1993 
to build a new apparatus, which started observations in June 1996. 
We present our new instrument and first results of some of its 
programs in this contribution. 
\section{The experiment}
The EROS-II\footnote{participating institutes : CEA-Saclay, IN2P3-CNRS, INSU-CNRS} 
instrument is a 1m diameter f/5 Ritchey-Chretien 
telescope, the MarLy, previously used in the french Alps until 
the mid-80s. It has been  
specially refurbished and automated in view of a microlensing 
survey. It is in operation at the European Southern Observatory at La Silla 
(Chile) since July 1996. EROS-II should be running until 2002.
\par
The optics includes a dichroic 
beam splitter allowing simultaneous observations in two wide 
pass-bands (a blue one, $V_{EROS}$ and a red one, $R_{EROS}$). 
The field of the instrument 
is observed in each band by a mosaic of $2\times4$ 
Loral $2k\times2k$ pixels thick CCDs. The mosaics cover  
$.7\deg(\alpha)\times 1.4\deg(\delta)$ (.6 arcsec per pixel). 
The median seeing (FWHM) is about 2. arcsec. 
CCDs from each mosaic are readout in parallel by DSPs. The total readout time 
is 50s. 
The data acquisition is controlled by two 
VME crates (one per camera) which transfer images to Alpha 
stations where flat-fielding, quality check and 
archiving onto DLTs are performed. 
Images are 
finally sent for analysis to the CCIN2P3 in Lyons. 
We are also developing an alert capability 
by monitoring on site, the day following the observation, 
a sample of stable stars from the microlensing fields. 
This effort is presently limited by the building of the 
stable stars database ; up to now 5 alerts have been announced. 
\par
EROS-II is primarily aimed at the search for microlensing events. 
We do this in several directions : 
the Magellanic Clouds 
(60 fields for the Large, 10 for the Small), 
the Galactic Center (67 fields) and 4 areas within the 
Galactic Plane ($\approx$ 6 fields each). 
We are currently giving the highest priority to the 
microlensing search in new 
lines of sight (the Small Magellanic Cloud and the Galactic Plane), 
from which we present some preliminary results.
We also address other cosmologically important programs, 
such as a systematic study of Cepheids, 
a search for high proper motion stars 
and a semi-automated supernov\ae\ search.   
These programs  use images from different regions in the sky ; 
each night the schedule optimizes the observational conditions 
for each program.
\begin{figure}[tb]
\begin{center}
\plotfiddle{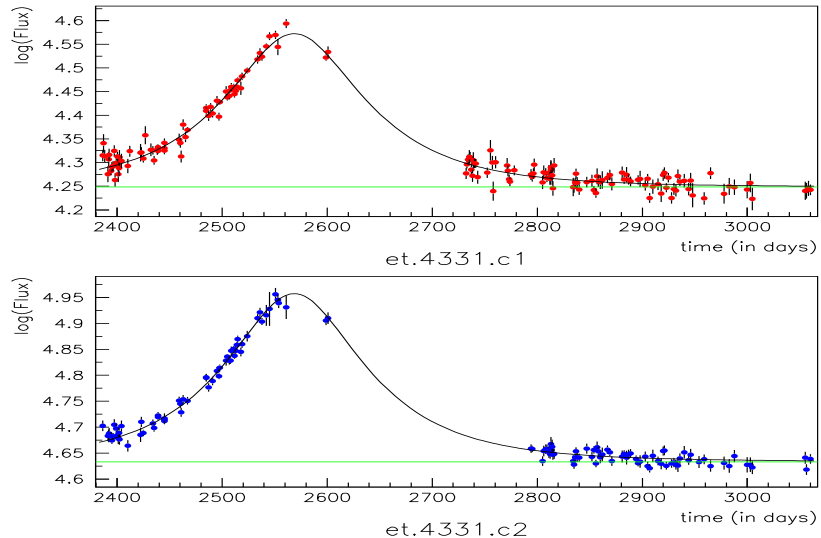}{7. true cm}{0}{60}{40}{-170}{0}
\caption{\label{smcand}The SMC-97-1 candidate light-curve with a 
standard microlensing fit superimposed 
(no blending assumed). Our time origin is Jan 0,1990.}
\end{center}
\end{figure}
\section{Microlensing towards SMC}
\subsection{The 1st year analysis} 
The analysis of the SMC data from our first running year is 
published in Palanque-Delabrouille 1998. 
The SMC was observed from July 1996 to February 1997 and then 
after July 1997 (our analysis includes data up to August 1997). 
A hundred images of each field are usable for subsequent analysis. 
In total more than $5.10^6$ light-\-cur\-ves could be analyzed.
They were sear\-ched for microlensing events. This 
selection is described in Palanque-Delabrouille 1998. 
The cuts are designed to select light curves with 
a unique and achromatic magnification, a suf\-fi\-cient S/N ratio and no known variable star 
contamination. The global selection efficiency 
is about 15\%. 
In our data 10 light-\-cur\-ves pas\-sed all cuts and where checked visually, 
one being finally selected : it 
is shown on figure \ref{smcand}. 
This event has also been seen by the Macho Colla\-boration (Alcock 1997). 
The amplified star is a blend of two stars  
with flux ratios of 70\% and 30\% ,
only the brightest of these two stars is amplified.
EROS also reported a periodic modulation of the combined flux with a 
period of 5.128 days and an amplitude of about 3\% 
of the brightest flux.\par
Using a simulation to compute our efficiency we also reported a first measurement of 
the Galactic halo optical depth $\tau$ 
towards SMC:
$\tau \simeq 3.3\ 10^{-7}$. 
Comparisons with several halo models show that this sole event con\-tri\-butes 
by about 40\% of the optical depth due to the halo of our Galaxy.
The event duration implies a most probable halo deflector mass of 
$2.6^{+8.2}_{-2.3}M_{\odot}$, which would make it improbable as a brown dwarf. 
The absence of parallax effect 
in this long duration event tends to imply either a heavy halo lens  
(a few $M_{\odot}$) or a light ($\leq .1 M_{\odot}$) deflector near the source. 
In the latter case, we have derived  
$\tau_{SMC-SMC} \simeq 1.3\ 10^{-7}$ with basic assumptions 
on the SMC struc\-ture (Palanque-Delabrouille 1998).
\begin{figure}[ht]
\begin{center}
\plotfiddle{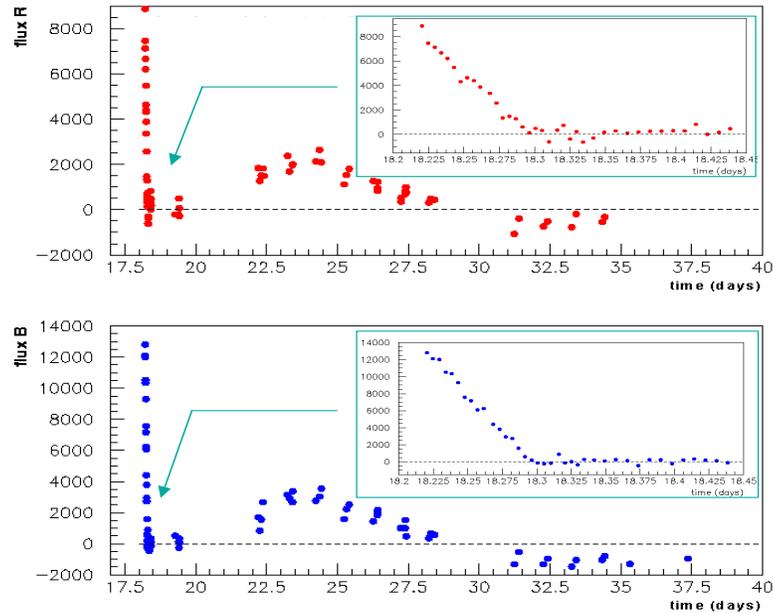}{7.5 true cm}{0} {65}{55}{-190}{-90}
\caption{\label{smcaus}EROS-II observations from the SMC-98-1 event,
analyzed using a differential photometry algorithm. 
We expanded the first day, showing the end 
of the caustic crossing (the time origin is Jan 0,1990).}
\end{center}
\end{figure}
\subsection{The SMC-98-1 Binary event}
On $25$ May 1998, the Macho collaboration sent 
an alert towards SMC. At that time, EROS-II was shutting down for a planned 
technical maintenance
(May $26^{th}$ - June $17^{th}$).  
After their observation on June $8^{th}$ of a dramatic increase in luminosity, 
interpreted as caused by a caustic crossing, it 
became clear that this event was induced by a binary lens. 
A second caustic crossing was predicted to occur around Jun $18^{th}$ by several groups,
including ourselves (using public data from the PLANET group). 
Due to the importance of this event, a high fraction of our observing time was 
immediately dedicated to observe this event during the first days after we restarted. As shown 
on figure \ref{smcaus}, we were lucky to observe in great details the end of the 2nd 
caustic crossing. 
Using this data alone, we could extract a limit on the caustic crossing time, which 
together with public data from MACHO enabled us to already put interesting 
constraints on the lens location, indicating only a 10\% 
likelihood for the deflector being a halo object (Afonso 1998). 
This was soon confirmed by other groups 
with much more confidence (see e.g. Bennett 1999 and references therein).
\subsection{Perspectives}
EROS-II has up to now observed two microlensing events towards SMC. 
We have obtained firm indications that the deflector causing them are 
both lying inside the SMC. The possibility for self microlensing inside the Magellanic 
Clouds receives therefore a strong support, in the case of the SMC. 
We are still continuing the data taking towards both Clouds. The analysis 
of data from our 2nd year towards SMC is near completion, and of course we are 
pushing hard the analysis of our LMC data
(more than 5 times more data than SMC). Within the next year, we should be able
to do a direct comparison between the LMC and SMC optical depths, which 
is important both for constraining Galactic halo models 
and elucidating the nature of the lenses.
\section{Microlensing towards the Galactic Plane}
\begin{figure}
\begin{center}
\plotone{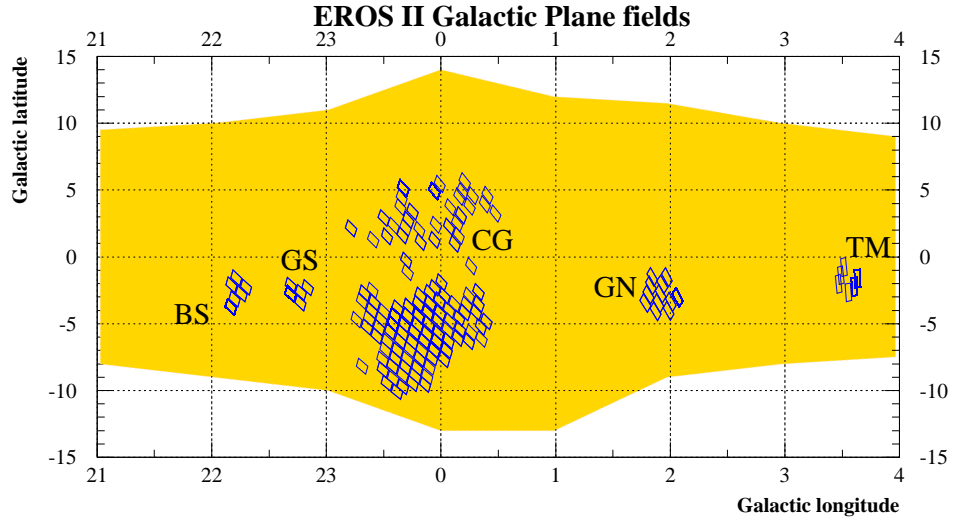}
\vspace{-.7 true cm}
\caption{\label{galmap}Map of our Galactic fields (galactic  coordinates, 
latitude is in degree and longitude in hour). 
The shaded area represents 
the shape of the Galaxy. We have indicated our Galactic Center fields (CG) and 
the 4 zones searched for microlensing outside the Center (GS \& BS - $\gamma$ and $\beta$ 
Scuti , GN - $\gamma$ Norm\ae\  and TM - $\theta$ Musc\ae ).} 
\end{center}
\end{figure}

Measuring optical depths in various directions in the Galactic Plane would 
help constraining the different Galactic components 
to the Bulge and LMC or SMC optical depths.
We chose several directions to look at, 
grouped respectively near 
galactic longitudes of $25^o$ (5 fields), $30^o$ (6 fields), $310^o$ (6 fields) and $320^0$ 
(12 fields). They are indicated in figure \ref{galmap}.
Exposures lie between 2 and 3 min, and are optimized to maximize the number of fields
in a given observing time, while not reducing the precision of each measurement 
(see Mansoux 1997).
The first images taken in  these directions showed that 
we were able to monitor about 10 million stars in these directions, 50\% 
of these with a photometric precision better than \mbox{10\%.}
The color-magnitudes from our fields 
were analyzed qualitatively with 
the help of a simulation of the evolution of a star population. 
They are found in rough agreement with what is expected from  
$10^8$ year old stars  
located at $6$ to $8kpc$ (within 10\% 
) and a reasonable reddening (Mansoux 1997).  
Sources lying in such a rather small distance ran\-ge is essential for understanding 
microlensing events.
\par
We have analyzed our data towards each 4 directions, up to the beginning of 1998, 
which correspond to an average of 90 measurements per star. 
The light curves were searched for microlensing using a similar selection than that used 
in the SMC analysis. 
This analysis, summarized on table \ref{taba}, produced 3 candidates (Derue 1999).
\begin{table}[h]
\begin{center}
\begin{tabular}{|c|c c|c|c|}
\hline
Direction & \multicolumn{2}{|c|}{Scutum} & Norma & Musca \\
Zones & BS & GS & GN & TM \\
\hline 
Fields & 6 & 5 & 12 & 6 \\
Measurements & 80 & 80 & 100 & 90 \\
\# stars & $1.7\ 10^6$& $1.5\ 10^6$ & $2.3\ 10^6$& $1.6\ 10^6$ \\
Distance (kpc) & $6.5\pm0.8$(?) & $6.5\pm0.8$ & $8.0\pm0.6$(?) & ? \\
\hline
\hline
Candidates & 1 & 0 & 2 & 0 \\
Durations (days)  & 73 & - & 98\& 72 & - \\
\hline
\hline
Optical depth & \multicolumn{2}{|c|}{$0.5\ 10^{-6}$} & $0.56\ 10^{-6}$ & - \\
\hline
\end{tabular}
\caption{\label{taba} Summary of the microlensing search in the Galactic Plane} 
\end{center}
\end{table}
 \begin{figure}
 \begin{center}
 \plottwo{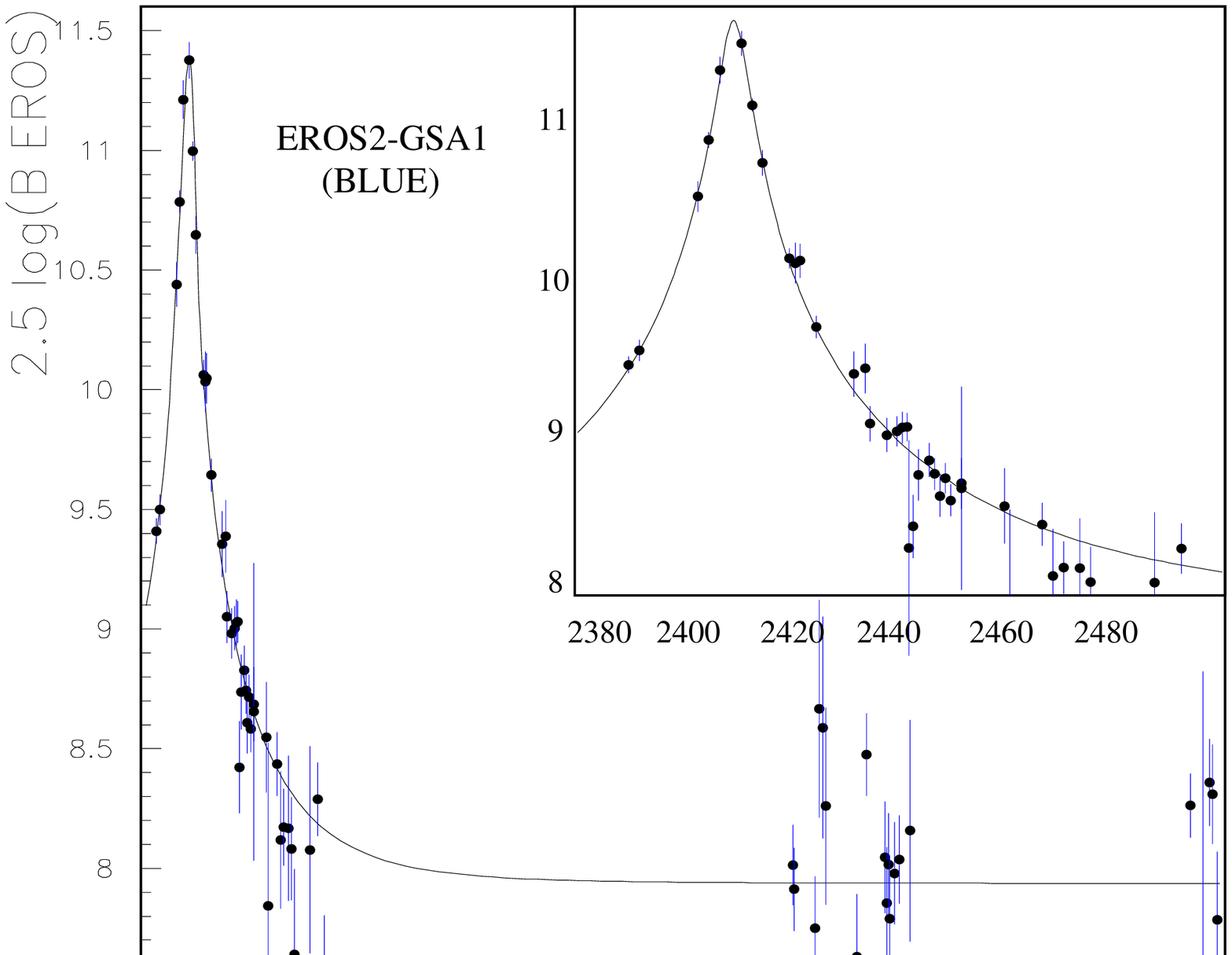}{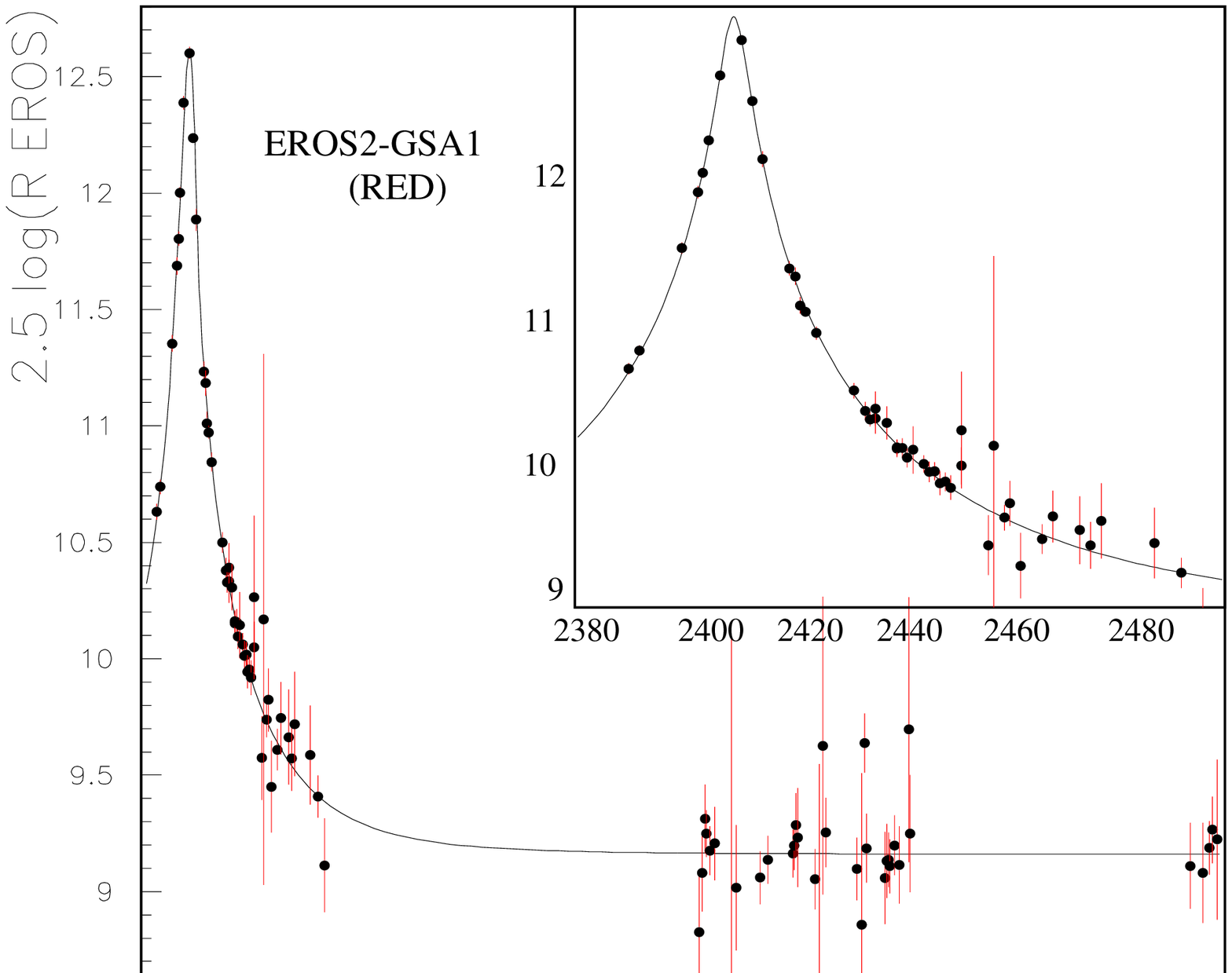}
 \vspace{4mm}
 \caption{\label{gsa1}GSA-1 light curve. We have superimposed 
on the data a fitted microlensing curve. Note the high magnification.}
 \end{center}
 \end{figure}
\begin{figure}
 \begin{center}
 \plottwo{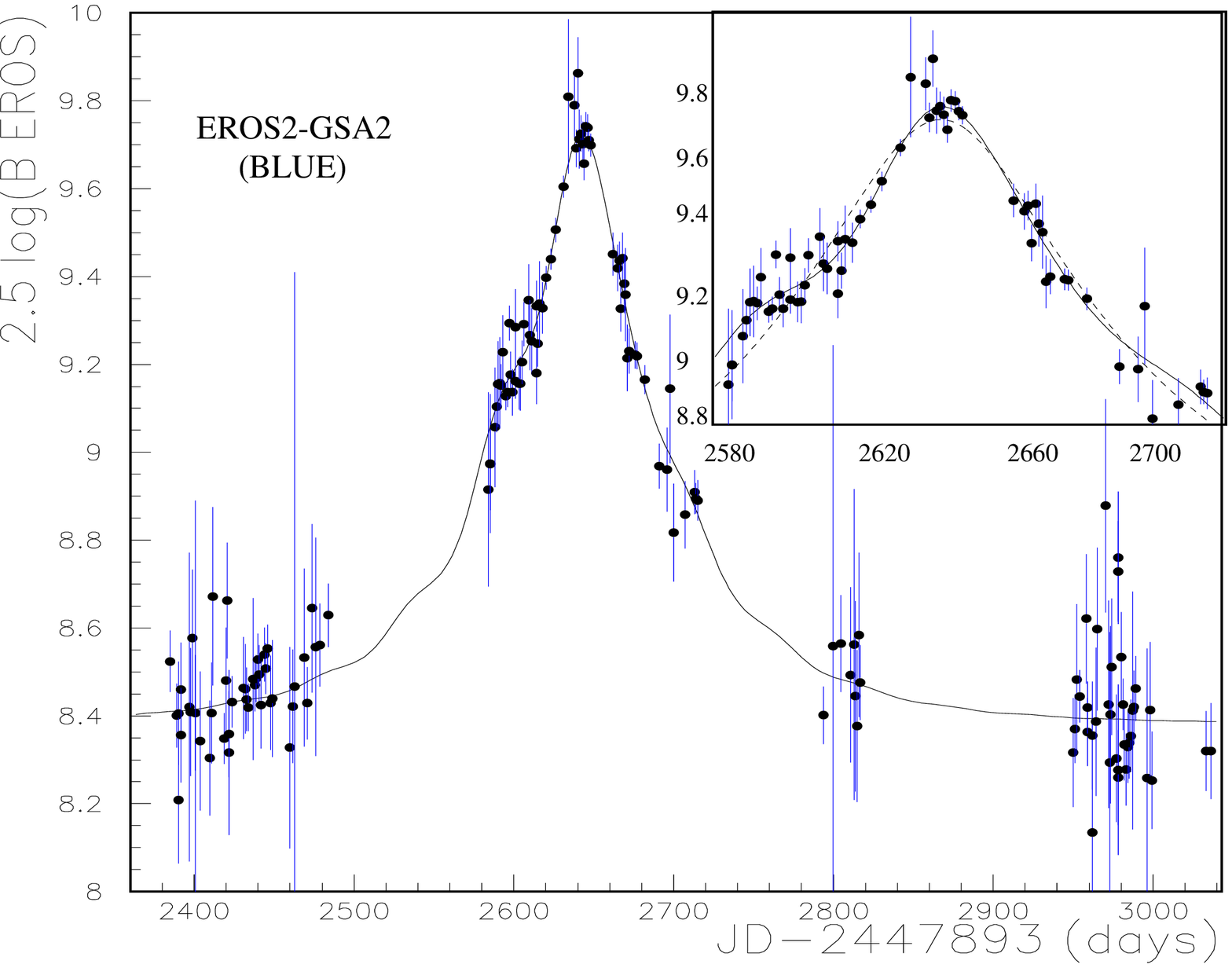}{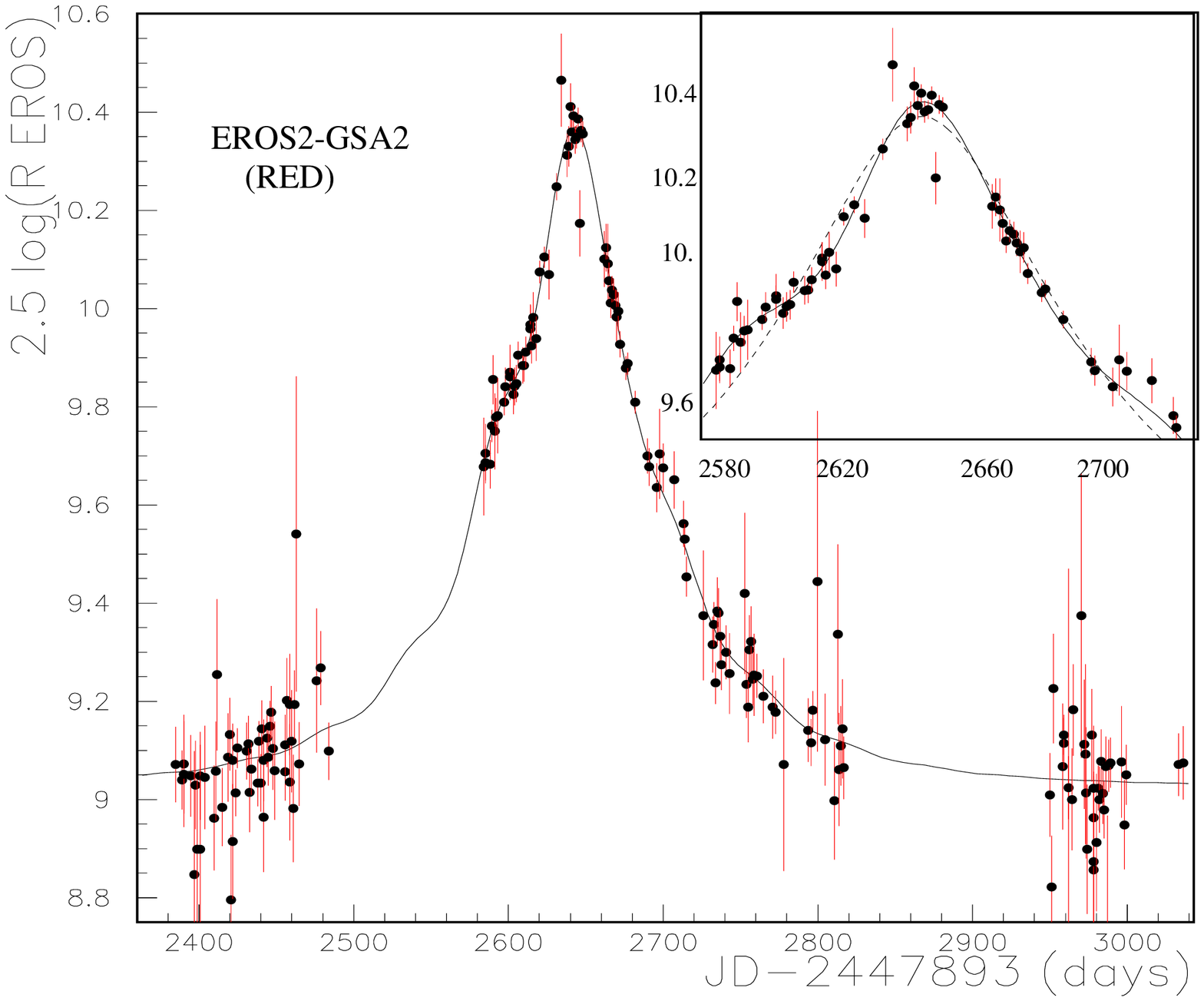}
\vspace{-2 mm}
 \caption{\label{gsa2}GSA-2 light curve. Superimposed onto the data point is 
a fitted curve for a binary source (50 days orbital period). In the zoom around the peak 
one may compare with the best fit for a static source.} 
 \end{center}
 \end{figure}
\begin{figure}
 \begin{center}
 \plottwo{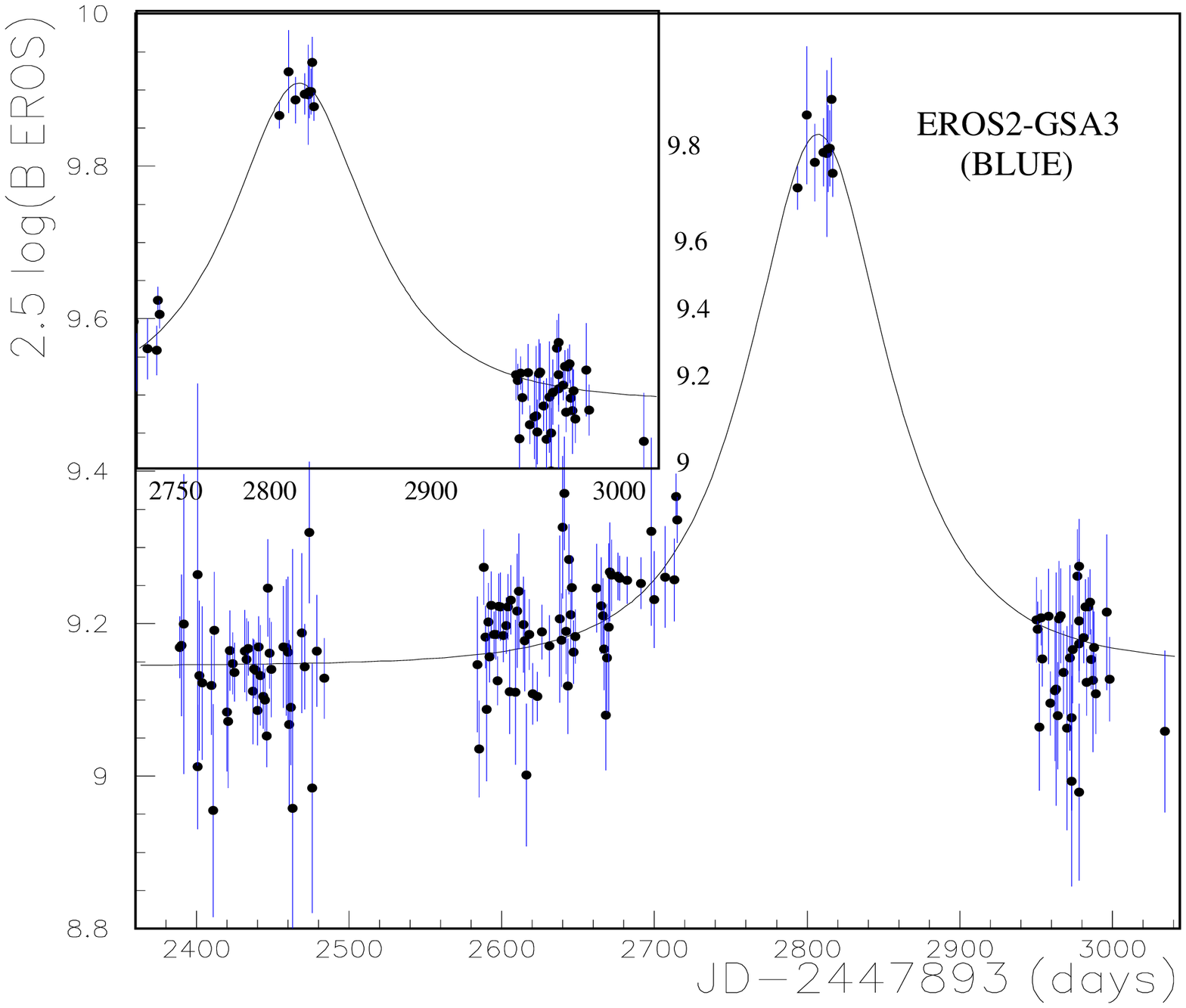}{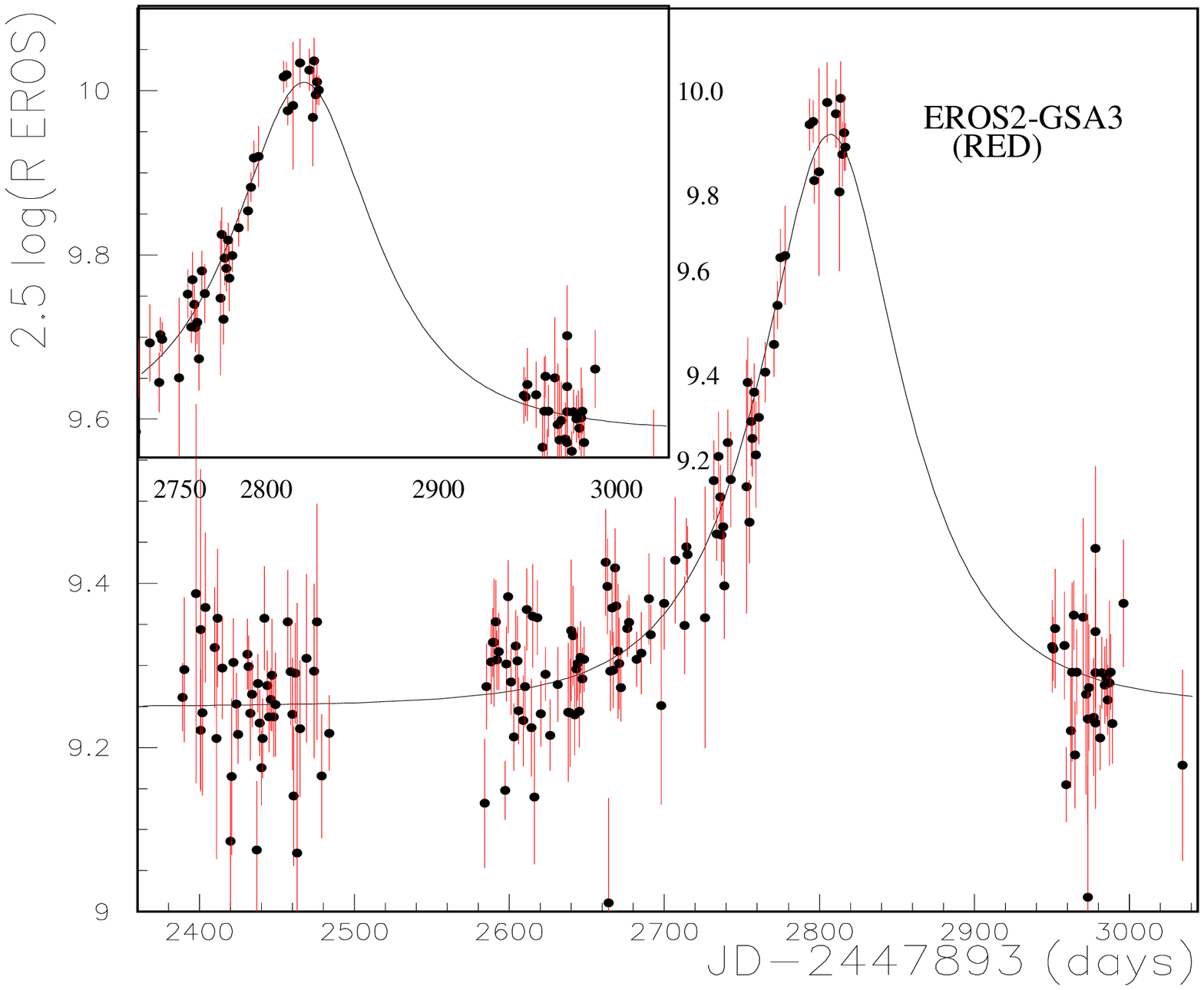}
\vspace{-2 mm}
 \caption{\label{gsa3}GSA-3 light curve. Superimposed onto the data point is 
a fitted standard microlensing curve.}
 \end{center}
 \end{figure}
Their duration is on average longer than those of the detected Bulge 
events (80 and 30 days respectively). This may be explained as follows. For the directions 
examined here, the lenses are located in the disk and have therefore 
a slow transverse motion ($\approx 40\ kms^{-1}$ - smaller than the disk/bulge relative transverse
velocity). 
The light curves of our candidates are shown on figure \ref{gsa1} to \ref{gsa3}.
These two events are peculiar. GSA-1 is a high amplification event, but neither blending nor 
parallax effects are seen. GSA-2 presents non-standard features, 
which are most probably explained by assuming that the source star lies in a orbiting binary 
system. This orbital motion modulates the line-of-sight (the ``xallarap'' effect). 
We have two satisfactory fits to the data, one with a period of $53\pm5$ days and 
one with $95\pm10$ days. 
We have computed the observed optical depths using preliminary efficiency 
calculation, uncorrected for blending (for which a better knowledge of the stellar population
is required). 
These roughly agree with our expectations from 
simple Galactic models (Mansoux 1997). 
After these promising results, these directions 
are still monitored, in order to increase our statistics. We did 
also some observations in order to improve our knowledge of the source stars 
(nature, distance). We hope within the duration of the experiment to 
be able to extract from these directions valuable informations 
which would help understanding Galactic dynamics as well as 
microlensing results towards the other directions (the Bulge and the Clouds).
\section{Magellanic cepheids} 
EROS-I reported a possible 
metallicity effect on the period-luminosity-color relation 
for SMC and LMC cepheids (Sasselov 1997). 
We pursued in 1996-7 a specific program 
in order to study this effect more accurately. Two fields  
in the center of LMC and two in that of SMC were imaged each night. 
These fields partially overlap those covered by the EROS-I CCD experiment.
At the end, 110-160 images per field entered the analysis. 
The photometric reduction followed the same path as our microlensing images. 
The resulting light curves were  
searched for periodic variations. Cepheids were 
subsequently isolated using cuts in the C-M and Period-Luminosity diagrams, 
and finally a visual selection. 
Our new sample contains about 300 cepheids in 
LMC and 600 in SMC (Bauer 1997) to be compared with 80 and 400 respectively
in the EROS-I database.  
Being observed with the same instrument and in similar atmospheric conditions 
also reduces systematics. This large catalog will soon be published.
\par 
\begin{figure}
\begin{center}
\plotfiddle{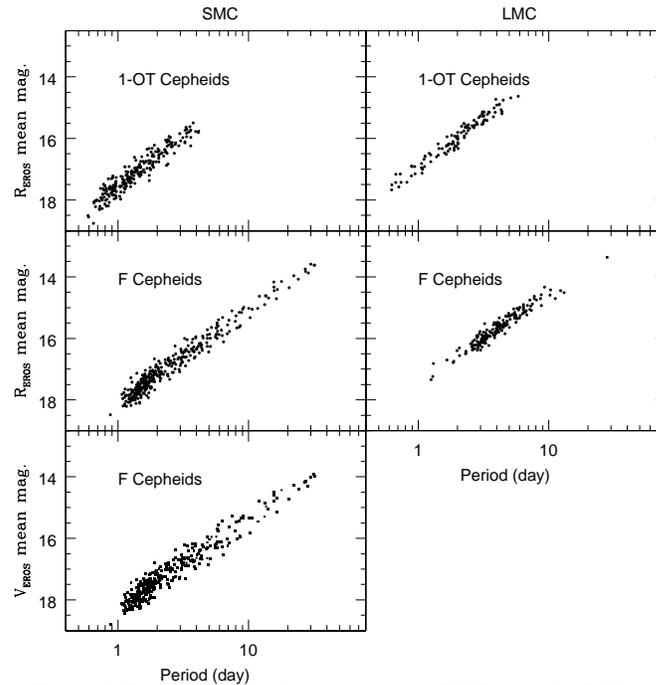}{7.5 cm}{0}{47}{47}{-130}{-90}
\caption{\label{plcep}
Period-Luminosity diagrams for LMC and SMC cepheids, in the $R_{eros}$ band. 
The P-L diagram in $V_{eros}$ is also shown for F-cepheid, where the non-linearity 
at low period is seen.}
\end{center}
\end{figure}
Some systematic studies using this large catalog have been performed. 
When building the Period-Luminosity diagrams,  
for short period cepheids, typically $T\geq2.5\ days$, a non-linearity exists  
in both colors (Bauer 1998). We did systematic checks of this effect, 
seen significantly for the first time. Among them, we 
checked that it was already 
present in the EROS-I data although not statistically significant. 
This rules out much of the possible observational biases. 
This effect is in fact accounted for by the evolutionary models 
developped by Baraffe 1998. One should also 
remark cepheids used in the measurement of extragalactic distances 
have periods above this non-linearity. 
\par
This study will of course be extended to the sample of cepheids 
isolated within our microlensing data. We are also 
analyzing the sources of the observed dispersion in the P-L diagrams, 
searching for effects like a spread in distance of the cepheids (Graff 1998).
\section{The supernov\ae\ search}
EROS-II has also started a semi-auto\-mated supernov\ae\ search. It is aimed at 
discovering in a programmed way batches of supernov\ae . 
Spectrography and photometry observations may thus be planned   
in order to classify and study them accurately. 
Supernov\ae\ are rare phenomena ($\approx$ 1 per century and per galaxy) 
and stu\-dy a large ($\approx 100$) number 
of them offers interesting cosmological pers\-pec\-tives, for example measuring their ra\-te, 
or study type Ia SN. 
Systematic sur\-veys for nearby SNe have 
tackled to their usability as distance indicators  
(with  $\approx$ 20 SNe, Hamuy 1996). This 
could be well studied in an intermediate redshift 
search such as that of EROS. Our detection threshold is estimated to be 
about the 22nd magnitude in V (10 mn exposures), corresponding to a redshift of 0.2. 
Such systematic studies are also essential in extracting cosmological 
informations from the distant SNe searches (Kim 1997, Leibundgut 1997).
\par
\begin{figure}
\begin{center}
\plotone{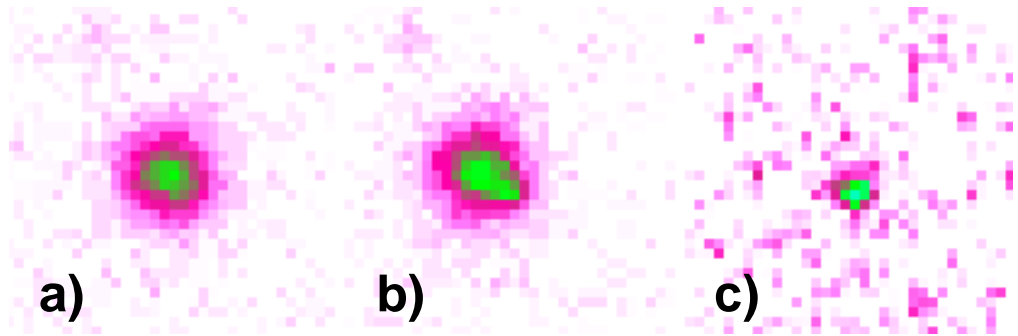}
\caption{\label{sn}Pictorial view of the discovery of SN1997bl. The image (a) 
is our reference image, taken on 03/02/97, (b) is the discovery image  
taken on 07/03/97, showing the SN superimposed onto the host galaxy.
(c) is the difference between (b) and (a), on which the supernova was detected.}
\end{center}
\end{figure}

Our program goes as follows. Using the EROS-II instrument, we take 
images near two new moons. The more recent (current) image is compared 
to the older one (reference) during daytime to search for SN, as 
shown on figure \ref{sn}. 
This program was tested in spring 1997. Three supernov\ae\ 
were discovered, in good agreement with our estimated discovery rate  
of $.1\ SN.\hbox{deg}^{-2}$. More intensive 
campaigns followed in the end of 1997 and in 1998. We obtained 
photometric follow-up time on the ESO 1.5m Danish telescope, and have also 
few spectroscopic nights on the ARC 3.5m telescope (New-Mexico, USA), 
on the WHT telescope (Canaries) and on the 3.6m ESO telescope. 
28 SNes were discovered during this period, at rate of about 
4-5/new moon (we search for about 50deg$^2$ each new moon, when 
weather permits). Ten of them have been  
identified through a spectrum.
\par
During this first intensive campaign, we were 
able to measure the SN explosion rate (Hardin 1998).
We plan to intensify our follow-up efforts (contributions are 
welcome). We should also participate to a ``joint nearby search'' 
coordinated by the SN Cosmology Project in February-March 1999. 
\section{Conclusion}
EROS-II started taking data more than two years ago.  
The instrument has been running quite well. We present a measurement 
of the galactic halo microlensing optical depth towards SMC based on one event
found in the 1996-7 data. 
This event could however as well be interpreted as due to a lens 
lying within SMC, and presents some intriguing properties. 
A second event was triggered in 1998, which was a binary event with caustic 
crossings. Our follow-up data indicates that the lens is also probably within the SMC.
The results from the analysis of our 2nd year towards SMC and 
from our LMC data should therefore bring some light on the nature of the lenses.
We also conducted a microlensing search towards directions in the 
galactic plane, and isolated the first 3 candidates in these 
directions. Precise measurements of the optical depths should 
constraint Galactic models. 
We are also analyzing a large LMC and SMC 
cepheids database to check 
for systematic effects on the distance scale deduced from them.  
We have found a new feature in the P-L diagram which is non-linear 
for short period($T\leq 2.5\hbox{days}$) SMC F-Cepheids. 
Our SNe search has been quite intense between fall 1997 and the summer of 1998, 
leading to the discovery of $\approx$ 30 SNe that have been scarcely followed-up 
with our instrument and others. These subjects are only a subsample 
of the rich physics outcome that is to be expected soon from EROS-II.

\end{document}